\documentstyle[12pt,amssymb]{article}
\catcode`@=11
\newcount\@tempcntc
\def\@citex[#1]#2{\if@filesw\immediate\write\@auxout{\string\citation{#2}}\fi
  \@tempcnta\z@\@tempcntb\m@ne\def\@citea{}\@cite{\@for\@citeb:=#2\do
    {\@ifundefined
       {b@\@citeb}{\@citeo\@tempcntb\m@ne\@citea\def\@citea{,}{\bf ?}\@warning
       {Citation `\@citeb' on page \thepage \space undefined}}%
    {\setbox\z@\hbox{\global\@tempcntc0\csname b@\@citeb\endcsname\relax}%
     \ifnum\@tempcntc=\z@ \@citeo\@tempcntb\m@ne
       \@citea\def\@citea{,}\hbox{\csname b@\@citeb\endcsname}%
     \else
      \advance\@tempcntb\@ne
      \ifnum\@tempcntb=\@tempcntc
      \else\advance\@tempcntb\m@ne\@citeo
      \@tempcnta\@tempcntc\@tempcntb\@tempcntc\fi\fi}}\@citeo}{#1}}
\def\@citeo{\ifnum\@tempcnta>\@tempcntb\else\@citea\def\@citea{,}%
  \ifnum\@tempcnta=\@tempcntb\the\@tempcnta\else
   {\advance\@tempcnta\@ne\ifnum\@tempcnta=\@tempcntb \else \def\@citea{--}\fi
    \advance\@tempcnta\m@ne\the\@tempcnta\@citea\the\@tempcntb}\fi\fi}
\catcode`@=12

\begin{document}

\thispagestyle{empty}
\hfill\\[10mm]
\begin{center}
{\bf\large New Insight into the Relation between
Torsion and Electromagnetism}\\[40mm]
{ Kenichi Horie\footnote{E-mail: horie@dipmza.physik.uni-mainz.de}}\\[10mm]
{\it Institut f\"ur Physik, Johannes Gutenberg--Universit\"at,
D--55099 Mainz, Germany }\\[2mm]
( March 23, 1995 )\\[10mm]
\end{center}
\begin{flushleft}{\bf Abstract}\\[2mm]\end{flushleft}
In several unified field theories the torsion trace is set equal to the
electromagnetic potential. Using fibre bundle techniques we show that this
is no leading principle but a formal consequence of another geometric
relation between space-time and electromagentism.

\newpage

Torsion in general relativity is commonly studied within the framework of
Einstein--Cartan theory, in which it is related to spin \cite{heh76}. However,
there is also another physical role of torsion suggested in several
works on the unification of gravity and electromagnetism
\cite{bor,mck,jak,fer}. The idea of such a geometric
unification is to omit any restrictions on the linear connection $\Gamma^
\alpha{}_{\!\mu\beta}$ and to identify its torsion trace
$T_\mu=T^\alpha{}_{\!\mu\alpha}=\Gamma^\alpha{}_{\!\mu\alpha}-\Gamma^
\alpha{}_{\!\alpha\mu}$ with the electromagnetic potential $A_\mu$.

It is well-known that Einstein's so-called non-symmetric unified field
theory of gravity and electromagnetism \cite{ein} suffered from
severe inconsistencies. Subsequently, several authors tried to remedy these
drawbacks by changing the employed Lagrangian and by introducing the ansatz
$T_\mu\sim A_\mu$ in an ad hoc manner \cite{bor}. Later on, this ansatz
could be motivated by the structure of the field equations, which precisely
resembled the Einstein--Maxwell equations \cite{mck,jak}. Thereby, an
arbitrary connection $\Gamma^\alpha{}_{\!\mu\beta}$ was restricted by the
field equations to be of the form
\begin{equation}
  \Gamma^\alpha{}_{\!\mu\beta}=\big\{{}^\alpha{}_{\!\mu\beta}\big\}
                              +\frac{1}{3}\delta^\alpha{}_{\!\beta}T_\mu\;,
\end{equation}
where $\big\{{}^\alpha{}_{\!\mu\beta}\big\}$ is the Christoffel symbol.
Despite the formal agreement of the field equations, the proposed
identification
$T_\mu\sim A_\mu$ lacked a clear geometric and physical meaning,
because $T_\mu$ is only a vector but not an $\mbox{U}(1)$ potential
like $A_\mu$ and therefore can not be gauged. The so-called
$\lambda$--transformation, introduced first by Einstein in another context
\cite{ein}, could not substitute the $\mbox{U}(1)$ structure, since its
geometric foundation is obscure.

The real problem with the ansatz $T_\mu\sim A_\mu$ is that no true $\mbox{U}
(1)$ fibre bundle structure have been constructed. In \cite{fer} such a
structure was introduced, but it differed from the common understanding
of $\mbox{U}(1)$ gauge theory. For example, charged particles were
represented by scalar densities of an ``imaginary weight''. A related
problem with unified field theories is the lack of a physical
interpretation of the resulting connection (1): Since it is not metric,
$\nabla_{\!\mu}g_{\alpha\beta}=-\frac{2}{3}T_\mu\cdot g_{\alpha\beta}\neq 0$,
it must not be applied for the parallel transports of signals on the
space-time because this would lead to the dependence of physical invariants
upon their histories like in Weyl's unified theory \cite{wey}. Therefore,
it is necessary to decompose the whole connection (1)
into a metric part and the torsion term. But this can be done in several
ways, for example, as
\begin{equation}
  \Gamma^\alpha{}_{\!\mu\beta} =
  \big[ \big\{{}^\alpha{}_{\!\mu\beta}\big\} \big] +
  \big[ \frac{1}{3}\delta^\alpha{}_{\!\beta}T_\mu \big]
\end{equation}
or
\begin{equation}
  \Gamma^\alpha{}_{\!\mu\beta} =
  \big[ \big\{{}^\alpha{}_{\!\mu\beta}\big\}
       +\frac{1}{6}(\delta^\alpha{}_{\!\beta}T_\mu-T^\alpha g_{\mu\beta})
  \big] +
  \big[ \frac{1}{6}(\delta^\alpha{}_{\!\beta}T_\mu+T^\alpha g_{\mu\beta})
  \big] \;.
\end{equation}
In both examples the first bracket $[\ldots]$ represents a metric connection.
Although (2) is suggested by the field equations, there is no unique
geometric prescription to decide between the different choices of the
decompositions.

In principle, it is not difficult to provide a clear fibre bundle geometric
prescription of how to separate (1):  A general linear connection is
represented by a 1--form $\omega$ (with some characteristic features) on
the tangent frame bundle $F(M)$ of the space-time manifold $M$. Let $L$ be the
(special) Lorentz group and $L(M)$ the bundle of orthonormal frames.
Suppose now that $\omega$ can be definitely pulled back onto a
fibre-product bundle \cite{kob} of $L(M)$ and some $\mbox{U}(1)$ bundle
$\mbox{U}(1)(M)$. This is a principal bundle with structure group
$L\times\mbox{U}(1)$ and will be simply denoted by $(L\times\mbox{U}(1))(M)$.
Since this fibre-product bundle is built canonically from both bundles $L(M)$
and $\mbox{U}(1)(M)$, it is possible (see e.g.\ \cite{kob}) to decompose
$\omega$ uniquely into a metric connection 1--form on $L(M)$ and a potential
on $\mbox{U}(1)(M)$ and represent $\omega$ as the sum of these two
connection 1--forms. This would provide the desired separation
prescription of (1).

To make this pull-back idea more concrete, let us consider Dirac spinors $\psi$
\cite{hor}. It is well-known that spinor derivatives can be constructed not
only from the Christoffel symbol but also from any metric connection with
non-vanishing contorsion \cite{heh71}. By writing such a metric
connection in its orthonormal anholonomic components $\Gamma_{a\mu b}$ the
spinor derivative is defined by
\begin{equation}
\nabla_{\!\mu}\psi = \partial_\mu\psi
                    -\frac{1}{4}\Gamma_{a\mu b}\gamma^b\gamma^a\psi\;,
\end{equation}
where $\gamma^b\gamma^a$ have been employed instead of the commonly
used Lorentz generators $\sigma^{ba}=\frac{1}{2}(\gamma^b\gamma^a-
\gamma^a\gamma^b)$ in virtue of
metricity or, equivalently, the Lorentz algebra condition $\Gamma_{a\mu b}=
-\Gamma_{b\mu a}$ \cite{heh71,ber}. If we now omit this condition and use a
general linear connection instead, its non-vanishing trace part
$\Gamma_{a\mu b}\cdot\frac{1}{2}(\gamma^b\gamma^a+\gamma^a\gamma^b)=
\Gamma_{a\mu b}\eta^{ba}=\Gamma^a{}_{\!\mu a}$, $\eta^{ab}=\mbox{diag}
(+1,-1,-1,-1)$, also contributes to the spinor derivative,
\begin{equation}
\nabla_{\!\mu}\psi = \partial_\mu\psi
                    -\frac{1}{4}\Gamma_{a\mu b}\sigma^{ba}\psi
                    -\frac{1}{4}\Gamma^a{}_{\!\mu a}\psi\;.
\end{equation}
The merits of this extended spinor derivative are manifold: Already at this
formal level the connection is clearly decomposed in its metric part
$\frac{1}{2}(\Gamma_{a\mu b}-\Gamma_{b\mu a})$, and its non-metricity vector
$\frac{1}{4}\Gamma^a{}_{\!\mu a}$, which will become the $\mbox{U}(1)$
potential. The extension of the spinor derivative (4) is not unique
since $\sigma^{ba}$ could
have been replaced equally well by $-\gamma^a\gamma^b$ or, more generally,
by $\sigma^{ba}+\varepsilon\cdot\eta^{ba}$. Due to this freedom, spinors
with any multiple of the elementary charge, $\varepsilon e$, can be treated.
Another merit of (5) is that, besides the electromagnetic
phenomena, the spin-torsion coupling established in Einstein--Cartan
theory is automatically included. The most important consequence of (5) is,
that the field equations now enforce a complex rather than a real valued
connection. This complex extension is essential to the construction of the
correct $\mbox{U}(1)$ bundle structure \cite{hor}.

To explain the geometric content of (5) and, at the same time, to deduce the
required decomposition principle of (1), let us look first at the usual
spinor derivative (4): A metric connection 1--form $\omega_m$ is defined
on $L(M)$ only, which --- provided that $M$ is spin --- is endowed with a
spin structure $\mbox{Spin}(M)\rightarrow\!\!\!\rightarrow L(M)$. This is a
twofold covering bundle map and induces a ${\Bbb C}^4$ spinor bundle, on
which spinors with their spin 1/2 representation can be properly defined.
$\omega_m$ can be pulled back to $\mbox{Spin}(M)$ and yields a spin
connection, which in turn defines the spinor derivative (4). On the other
hand, a complex linear connection $\omega_c$ is defined on the whole complex
frame bundle $F_c(M)$, built from all tangent bases of ${\Bbb C}\otimes
TM$. Since there is no comparable twofold mapping onto $F_c(M)$,
$\omega_c$ does not yield a spin connection directly. Therefore, it must be
pulled back to an ``intermediate bundle'', for which a spin structure
exists. Such a bundle is given by $({\Bbb C}L\times\mbox{U}(1))(M)$,
which is the complex analogue of $(L\times\mbox{U}(1))(M)$ mentioned above
and is built from the
complexified orthonormal frame bundle ${\Bbb C}L(M)$ and a trivial
$\mbox{U}(1)$ bundle $M\times\mbox{U}(1)$. The fact that $\omega_c$ can
indeed be pulled back to this fibre-product, which in itself is not a natural
subbundle of the frame bundle, is not as trivial as it might first look
\cite{hor}. Once $\omega_c$ is pulled back onto this intermediate, a
complexified spin structure ${\Bbb C}\mbox{Spin}(M)\rightarrow\!\!\!
\rightarrow{\Bbb C}L(M)$ can be employed to further pull it back to
$({\Bbb C}\mbox{Spin}\times\mbox{U}(1))(M)$, which then gives rise to
the extended spinor derivative (5).

According to this geometric background, the linear connection
$\Gamma_{a\mu b}$ can be uniquely decomposed into its metric connection
$\frac{1}{2}(\Gamma_{a\mu b}-\Gamma_{b\mu a})$ on ${\Bbb C}L(M)$
and a true $\mbox{U}(1)$ potential $\frac{1}{4}\Gamma^a{}_{\!\mu a}$ on
$M\times\mbox{U}(1)$. In vacuum, if the linear connection is written in its
holonomic coordinate components $\Gamma^\alpha{}_{\!\mu\beta}$, the field
equations of the theory \cite{hor} yield the same result as in (1). But now
the above fibre bundle geometry unambiguously prescribes the decomposition
(2), see \cite{hor}.
The true geometric interpretation of electromagnetism is now given by
\begin{equation}
  \frac{1}{4}\Gamma^a{}_{\!\mu a} := \frac{ie}{\hbar c}A_\mu\;.
\end{equation}
Strictly speaking, $\frac{1}{4}\Gamma^a{}_{\!\mu a}$ is a 1--form defined on
the space-time manifold $M$, which has been obtained by pulling the
corresponding $\mbox{U}(1)$ potential on $M\times\mbox{U}(1)$ back onto
$M$ via a special $\mbox{U}(1)$ cross section (namely the trivial cross
section, which prescribes to each point on $M$ the constant value
$1\in\mbox{U}(1)$). If, instead, another $\mbox{U}(1)$ cross section is
used for the pull-back, then it will result in an $\mbox{U}(1)$ gauge
transformation of (6).
Now, the identification (6) can be inserted into the expression of the
whole connection (1), from which its torsion trace can be computed,
\begin{equation}
  T_\mu = 3\cdot\frac{ie}{\hbar c}A_\mu\;.
\end{equation}
Thus, $T_\mu$ still seems to be related to $A_\mu$. However, since the
coordinate connection components in (1) and also in (7) are obtained
by pulling back $\omega_c$ from the frame bundle to $M$ via the cross
section given by a coordinate reference frame ($\partial / \partial
x^\mu$), there is no possibility of an $\mbox{U}(1)$ gauge
transformation in (7). Therefore, to obtain (7) from (6), the special
$\mbox{U}(1)$ gauge implicitly chosen in (6) must be held fixed.
Since (7) is valid in this $\mbox{U}(1)$ gauge on $M\times\mbox{U}(1)$
only, the relation $T_\mu\sim A_\mu$ is merely a formal remnant of the
true $\mbox{U}(1)$ identity (6), see for more information \cite{hor}.

Contrary to the unified field theories \cite{bor,mck,jak,fer}, where
the whole connection (1) is
supposed to unify gravity, represented by the Christoffel symbol, and
electromagnetism, we have seen that the trace part $\frac{1}{4}
\Gamma^a{}_{\!\mu a}$ must be detached from the whole connection on the
frame bundle and pulled back to a $\mbox{U}(1)$ bundle in order to
obtain the electromagnetic
potential. This decomposition principle is in accord with the well-known
theorem that it is impossible to combine space-time and
internal symmetry in any but a trivial way \cite{ora}. We can say,
however, that it is not necessary to include the electromagnetic potential
into the space-time as something alien or, as has been done by
Infeld and van der Waerden \cite{inf}, only on the spin connection level,
but that electromagnetic phenomena can be viewed as originating from the
intrinsic geometry of space-time.  \\[3mm]
{\bf Acknowledgments}\\[1mm]
I thank Dr.\ P.\ O.\ Roll for reading the manuscript.

\end{document}